\documentstyle[11pt,epsfig]{article}
\begin{document}

\title{Extended Brinkman-Rice Picture and
Its Application to High-$T_c$ Superconductors\thanks{This is in
"New Trends in Superconductivity", P. 137 (2002), (Ed. J.F. Annett
and S. Kruchinin, Kluwer, Dordrecht).}}
\author{Hyun-Tak Kim\thanks{kimht45@hotmail.com.
htkim@etri.re.kr..}\\
\it Telecom. Basic Research Lab., ETRI, Taejon 305-350, Korea}
\date{}
\maketitle{}
\begin{abstract}
This is a full-paper of the paper(Physica C341-348, 259(2000))
published previously. The effective charge and Coulomb energy are
justified by means of measurement. Theoretical calculations of the
effective mass depending on band filling also are given by
Gutzwiller's variational calculations. On the basis of the concept
of measurement, the effective mass is an average of the true
effective mass in the Brinkman-Rice(BR) picture for metal phase
and is the effect of measurement. The true correlation strength
($U/U_c=\kappa_{BR}$) in the BR picture is evaluated as
0.92$<\kappa_{BR}<$1 for YBCO. High-$T_c$ superconductivity is
attributed to the true effective mass (regarded as the density of
states) caused by the large ${\kappa}_{BR}$ value. Furthermore,
the validity of the BR picture is indirectly proved through the
extended BR picture.\\
\end{abstract}

\section{Introduction}
{\hspace{0.6cm}}Several theoretical studies have been performed
to  reveal   the mechanism of the metal-insulator transition
(MIT) in 3$d$ transition-metal oxides  including strongly
correlated high-$T_c$ superconductors$[1-6]$. Mott first
introduced the concept of the Mott MIT of the first order (called
"Mott transition"), which is caused by a strong on-site Coulomb
repulsion $U [1]$. Hubbard$[2]$ made a model which explained the
splitting between the lower and upper Hubbard bands resulting in
increased $U$, which is indicative of an insulator and
second-order MIT. However, Hubbard's model has not clarified the
single-particle spectral function $\rho(\omega)$ characterizing
physical properties, particularly on the metallic side of the
Mott transition.

Brinkman and Rice$[3]$ took account of first order MIT from
results calculated using the Gutzwiller variational theory$[7]$
under the condition of equal numbers of spin-up electrons and
spin-down electrons, restricted to a metal with the electronic
structure of one electron per atom. They predicted that the
effective mass of a quasiparticle diverges and a MIT of first
order occurs at $U/U_c$=1, and that the width of the kinetic
energy decreases as $U/U_c{\rightarrow}1$. These results were
built on the Fermi liquid theory and have been called the
Brinkman-Rice (BR) picture. However, the BR picture does not give
exact information on the band-filling dependence of the effective
mass.

The $d=\infty$ Hubbard model suggested that the second-order MIT
occurs at $U/U_c$=1, and that the metallic side exhibits the BR
picture$[4]$. The one-dimensional(1-D) $t-J$ model$[5]$ and the
2-D Hubbard model$[6]$ accounted for the band-filling dependence
of the effective mass of a quasiparticle for strongly correlated
metals.

On the basis of these theories, many experiments have been
carried out to clarify MITs. Systematic experimental studies on
MITs and  the effective mass of quasiparticles have been made on
filling-controlled systems$[8-11,16,17]$. For
Sr$_{1-x}$La$_x$TiO$_3$,  the  sharp first-order MIT and the
critical effective  mass   were observed near $x$=0.95$[8]$.
Sr$_{1-x}$La$_x$TiO$_3$ seemed to follow  the BR picture, $t-J$,
and 2-D  Hubbard models   near the  Mott transition$[8,9]$.
However, the band-filling dependence of   the effective mass has
not been explained quantitatively$[8,9]$.

Morikawa et al.$[12]$, by analyzing the photoemission spectra,
suggested that metallic properties for CaVO$_3$ and SrVO$_3$
metals are governed by long-range interaction as well  as
short-range interaction. Inoue et al.$[13]$ and Makino et
al.$[14]$, by  controlling the bandwidth, reported that the
effective mass for Ca$_{1-x}$Sr$_x$VO$_3$ metals did not increase
critically near the Mott transition. These analyses differ from
those of models based on short-range interaction. One cause of
the difference is because the band-filling dependence of the
effective mass is left unspecified.

With regard to the high-$T_c$ superconductor
La$_{2-x}$Sr$_x$CuO$_4$, the parent material La$_2$CuO$_4$ has
been known to be an antiferromagnetic Mott insulator, although any
behavior indicative of the first-order transition has not been
observed near $x$=0. The effective mass of a quasiparticle
increases with increasing $x$ in the underdoped regime$[18]$.
This La-system seems to be explained by the Hubbard models rather
than the BR picture, $t-J$, or 2-D Hubbard models exhibiting
first-order transition. Up to now, there is no theory capable of
explaining the MITs of both Sr$_{1-x}$La$_x$TiO$_3$ and
La$_{2-x}$Sr$_x$CuO$_4$ systems.

On the generic question of high-$T_c$ superconductivity, since
discovering the high-$T_c$ superconductor, the van Hove
singularity (vHs), which is a 2D-density of states(DOS), has been
suggested as an explanation for the high-$T_c$ mechanism; the
2D-DOS is proportional to the effective mass of a quasiparticle.
However, through experimental results, many researchers have
pointed out that the vHs is unfit to account for high-$T_c$
superconductivity$[19-23]$. Instead, an extended vHs has been
observed by angle-resolved photoemission spectroscopy$[24,25]$.
This extended vHs may be a power law as a function of the
difference between the extended saddle-point energy and the Fermi
energy. The band-filling dependence of the 2D-DOS regarded as the
extended vHs still remains to be clarified by theory. The
mechanism of a pseudogap for the underdoped regime for
superconductors has long been discussed$[26-30]$. A new phase
diagram suggested that the pseudogap disappears at optimal doping
$[31]$.

In this paper, in order to reveal the above problems, first of
all, we review the paper which gives an insight regarding the
extension of the Brinkman-Rice picture and shows experimental
results$[32]$. The concept of measurement is introduced to
justify the effective Coulomb energy based on the metal-insulator
instability derived from the charge-density-wave(CDW) theory of
BaBiO$_3$. Theoretical calculations of the effective mass also are
given. Other examples not given in the paper, including data on
high-$T_c$ superconductivity and strongly correlated materials,
are appended, for the extended BR picture.

\begin{figure}[t]
\vspace{0.1cm}
\centerline{\epsfxsize=12.0cm\epsfbox{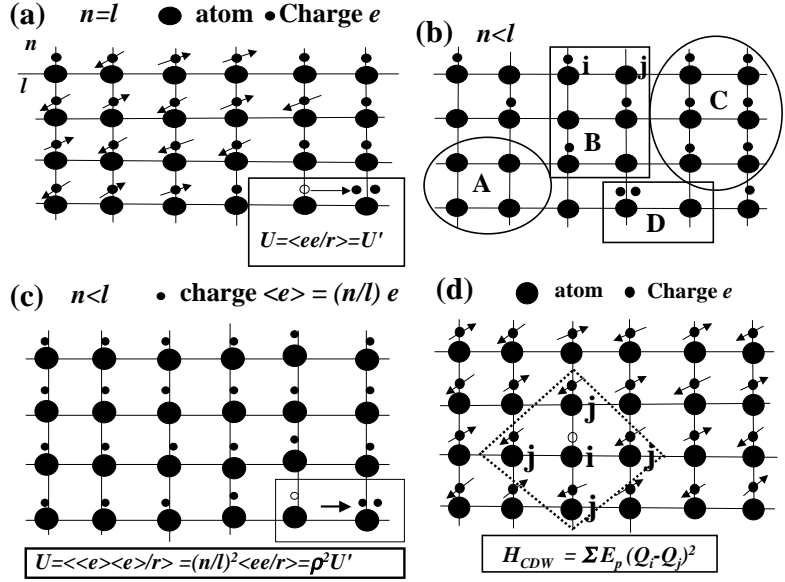}}
\vspace{0.1cm} \caption{\small{(a) In the case of one electron per
atom, the on-site Coulomb repulsion is given as
$U={\langle}\frac{e^2}{r}{\rangle}=U'$. (b) In the case of $n<l$,
the four possible electronic structures are region A (insulator),
region C (metal), and regions B and D (CDW insulators). Here, $n$
and $l$ denote the number of carriers in region C and the number
of atoms in the whole system, respectively. (c) In the case of
less than one electron per atom, the on-site Coulomb repulsion is
given as
$U=(n/l)^2{\langle}\frac{e^2}{r}{\rangle}={\rho}^2U^{\prime}$.
(d) In the case where one electron is removed, both an atom $i$
and the nearest neighbors $j$ of the atom become the CDW
insulator due to charge disproportionation, ${\delta}Q=Q_i-Q_j$.}}
\end{figure}

\section{Extended Brinkman-Rice picture}
\subsection{Justification of the effective Coulomb energy}
{\hspace{0.6cm}}A strongly correlated metallic system with a $s$
band structure is assumed for a $d=\infty$ dimensional
simple-cubic lattice. Let $n$ and $l$ be the number of electrons
(or carriers) and the number of atoms (or lattices),
respectively. In the case of one electron per atom, $i.e.$,
$n=l$, the metallic system is a metal and the existence
probability ($P=n/l={\rho}$= band filling) of electrons on
nearest-neighbor sites is one. The electrons have a spin on an
atom. The on-site Coulomb repulsion is always given by
$U=U'{\equiv}{\langle}{\frac{e^2}{r}}{\rangle}$, as shown in Fig.
1 (a). However, in the case of $n<l$ which occurs by doping an
element to a base insulator or metal, $U$ is determined by
probability. The metallic system is quite complicated, as shown
in Fig. 1 (b). Four types of regions for the system can be
distinguished as possible extreme examples. Region A in Fig. 1
(b) has no electrons on its atomic sites, which corresponds to a
normal insulator. Region C has a metallic structure of one
electron per atom. Regions B and D have a charge-density-wave
(CDW) structure unlike the assumed cubic lattice. Electrons in
region D have opposite spins. This must be regarded as a CDW
insulator with the CDW-energy gap depending on the local
CDW-potential energy$[33,34]$. Moreover, even when one electron
on an atom is removed, the nearest-neighbor sites of the atom
without an electron and the atom itself are regarded as region B,
as shown in Fig. 1(d). The potential energy can be interpreted as
a potential for impurities in the metallic system.

The CDW-potential energy, $V_{CDW}=-E_p(Q_i-Q_j)^2$, is derived by
breathing-mode distortion due to the charge disproportionation
(${\delta}Q=Q_i-Q_j$) between nearest-neighbor sites, where $Q_i$
and $Q_j$ are charges irrespective of spins on $i$ and $j$ sites,
respectively. The CDW gap is regarded as a pseudogap (or pinned
CDW gap), and can be observed by optical methods$[26-30]$,
photoemission spectroscopy$[27]$, and heat capacity
measurement$[35,36]$ for high-$T_c$ superconductors. The
pseudogap occurs if the number of carriers is less than the
number of atoms. Therefore, the metallic system is inhomogeneous
and cannot be self-consistently represented in $k$-space, $i.e.$,
$U$ and the density of states of the system are not given. To
overcome this difficulty, carriers on atoms in the metal phase
(region C) have to be averaged over sites, as shown in Fig. 1
(c). The on-site effective charge is
$Q_i=Q_j=<e>=(n/l)e={\rho}e$, on average over sites. The metallic
system, then, can act as a metal, because $V_{CDW}=0$ due to
${\delta}Q$=0. Therefore, the effective Coulomb energy is given by
$U={\rho}^2U^{\prime}$, as shown in Fig. 1 (c).

To illustrate the physical meaning of $P=(<e>/e)=\rho<1$, then
$\mu<<1$, $\rho=1-\mu<1$, where $\mu=1-\rho=n_b/l$~(or
$n_b=l-n$), $n_b$ is the number of bound charges bound by
$V_{CDW} [34]$. For the CDW insulating side, $\mu<<1$ indicates
that there are a few doubly occupied atoms, such as in region D.
For the metallic side, $\rho<1$ indicates that a number of
carriers exist. This suggests that region C is extremely wide.

The insulating and metallic sides correspond to two phases, which
are attributed to the metal-insulator instability (or instability
of the CDW-potential energy)$[34]$ at $\rho$=1 (half filling).
This indicates that a metal (perfect single phase) with the
electronic structure of Fig. 1 (a) is not synthesized. Junod et
al.$[37]$ suggested that even the best samples are not 100$\%$
superconducting, which supports the metal-insulator
instability$[34]$. Thus, synthetic metals composed of several
atoms always have the electronic structure of two phases, such as
Fig. 1 (b), which is a necessity of the development of
$<e>={\rho}e$.

To justify the fractional charge of carriers in the metallic
system with two phases, the concept of measurement is necessary.
When measuring the charge of carriers in the metallic system of
Fig. 1 (b), the measured charge becomes the effective fractional
one, $<e>={\rho}e$, representing the average meaning of the
system. This is because experimental data for the metallic system
are expectation values of statistical averages. When $not$
measuring charge, the charge of carriers in region C in Fig. 1
(b) remains the true elementary charge, not observed by means of
measurement.

Accordingly, for the metallic system where $n{\le}l$, the
effective fractional Coulomb energy, $U={\rho}^2U^{\prime}$, is
defined by using $<e>={\rho}e$ and justified by means of
measurement.

Additionally, although the justification is based on the metal
insulator instability derived from the CDW theory, it can be
applied to all strongly correlated metals including cuprate
superconductors. This is because there is much experimental
evidence for the CDW in cuprate superconductors although their
parent materials are antiferromagnetic insulators, which is shown
in a review paper$[38]$.

\subsection{Calculation of the effective mass}
{\hspace{0.6cm}}For the metallic system regarded as a real
synthetic metal with an electronic structure such as Fig. 1 (b),
the effective mass of quasiparticles needs to be calculated.
Hamiltonians of the metallic system can be considered as follows.
Hamiltonian, $H$, is given by

\begin{eqnarray}
H &=& H_1 + H_2,\\ H_1 &=&
\sum_{k}(a_{k\uparrow}^{\dagger}a_{k\uparrow}+
a_{k\downarrow}^{\dagger}a_{k\downarrow}){\epsilon}_k+ U
\sum_{g}a_{g\uparrow}^{\dagger}a_{g\downarrow}^{\dagger}a_{g\downarrow}a_{g\uparrow},
\\
H_2 &=& - \sum_{i,j}E_p(Q_i-Q_j)^2,
\end{eqnarray}
where $a_{k\uparrow}^{\dagger}$ and $a_{g\uparrow}^{\dagger}$ are
the creation operators for electrons in the Bloch state ${k}$ and
the Wannier state ${g}$, respectively, and ${\epsilon}_k$ is the
kinetic energy when $U$=0. $H_1$ and $H_2$ are Hamiltonians of the
metallic region C and the CDW-insulator regions B and D,
respectively. In the case of Fig. 1 (a) and (c), the Hamiltonian
is reduced to $H_1$ because $H_2$ disappears due to ${\delta}Q=0$,
and the on-site Coulomb energy is given by $U={\rho}^2U^{\prime}$.
$H_1$ is consistent with the Hamiltonian used in the Gutzwiller
variational theory$[7]$.

In order to calculate the effective mass of quasiparticles and the
ground-state energy for a strongly correlated metallic system, the
Gutzwiller variational theory$[7,39-41]$ is used. $H_1$ is
supposed to describe the metallic system. The wave function is
written as
\begin{eqnarray}
\vert\Psi\rangle={\eta}^{\bar\nu}{\vert\Psi}_0\rangle,
\end{eqnarray}
where ${\vert\Psi}_0\rangle$ is the wave function when $U=0$,
${\bar\nu}$ is the number of doubly occupied atoms, and
$0<{\eta}<1$ is variation. The expectation value of $H_1$ is
regarded to be
\begin{eqnarray}
{\langle}H{\rangle}=\frac{{\langle\Psi\vert\sum_{ij}\sum_{\sigma}t_{ij}a_{i\sigma}^{\dagger}a_{j\sigma}
 \vert\Psi\rangle}+{{\langle\Psi\vert}U{\sum_i\rho_{i\uparrow}\rho_{i\downarrow}\vert\Psi\rangle}}}
 {\langle\Psi\vert\Psi\rangle}.
\end{eqnarray}
The second part of the equation is simply given by
${{\langle\Psi\vert}U{\sum_i\rho_{i\uparrow}\rho_{i\downarrow}\vert\Psi\rangle}}$=$U\bar\nu$
because ${\vert\Psi}_0\rangle$ is an eigenstate of the number
operator ${\sum_i\rho_{i\uparrow}\rho_{i\downarrow}}$. The first
part is dealt with by assuming that the motion of the up-spin
electrons is essentially independent of the behavior of the
down-spin particles (and $vice ~versa$). By minimization with
respect to ${\eta}$, Gutzwiller obtained an extremely simple
result for the ground-state energy, namely,
\begin{eqnarray}
E_g/l=q_{\uparrow}({\bar\nu},\rho_{i\uparrow},\rho_{i\downarrow}){\bar\epsilon_{\uparrow}}+q_{\downarrow}
({\bar\nu},\rho_{i\uparrow},\rho_{i\downarrow}){\bar\epsilon_{\downarrow}}+U{\bar\nu}.
\end{eqnarray}
Here,
\begin{eqnarray}
{\bar\epsilon_{\sigma}}=l^{-1}{\langle\Psi\vert\sum_{ij}t_{ij}a_{i\sigma}^{\dagger}a_{j\sigma}
 \vert\Psi\rangle}=\Sigma_{k<k_F}\epsilon_k<0
\end{eqnarray}
is the average energy of the $\sigma$ spins without correlation
and $\epsilon_k$ is the kinetic energy in $H_1$, with the zero of
energy chosen so that $\Sigma_k\epsilon_k$=0.
${\bar\epsilon_{\uparrow}}$ is equal to
${\bar\epsilon_{\downarrow}}$.

The discontinuities, $q_{\sigma}$, in the single-particle
occupation number at the Fermi surface are given by
\begin{eqnarray}
q_{\sigma}=\frac{\left(\sqrt{(\rho_{\sigma}-{\bar{\nu}})(1-\rho_{\sigma}-\rho_{-\sigma}+
{\bar\nu})}+\sqrt{(\rho_{-\sigma}-{\bar\nu}){\bar\nu}}\right)^2}{\rho_{\sigma}(1-\rho_{\sigma})},
\end{eqnarray}
where $\rho_{\sigma}=\frac{1}{2}\rho$, $0<{\rho\le}1$ and
$\rho_{\uparrow}=\rho_{\downarrow}$ $[39,40]$. This, calculated by
Ogawa et. al.$[39]$ who simplified the Gutzwiller variational
theory, is in the context of the Gutzwiller variational theory.
Eq. (8) is a function of $\rho_{\sigma}$ and $\bar{\nu}$
irrespective of the quantity of charges. This can be analyzed in
two cases of $\rho$=1 and $0<\rho<$1, because Gutzwiller did not
limit the number of electrons on an atom for the metallic system.

In the case of $\rho$=1,
\begin{eqnarray}
q_{\sigma}=8\bar\nu(1-2\bar\nu).
\end{eqnarray}
This was described in the BR picture.

In the case of $0<\rho<1$, two kinds of $q_{\sigma}$ can be
considered. One is Eq. (8) when the electronic structure is
${\delta}Q\ne$0, as shown in Fig. 1 (b). However, Eq. (8) can not
be applied to the metallic system, as mentioned in the above
section. The other is Eq. (9) when the electronic structure is
${\delta}Q$=0, as shown in Fig. 1 (c). Eq. (9) is obtained from
substituting ${\rho}_{\sigma}$ in Eq. (8) with
${\rho}^{\prime}_{\sigma}$. Here,
${\rho}^{\prime}=(n^{\prime}/l)=1$ and
${\rho}^{\prime}_{\sigma}=\frac{1}{2}$, because the number of the
effective charges, $n^{\prime}$, is equal to $l$. It should be
noted that the metallic system with less than one electron per
atom, as shown in Fig. 1 (c), is mathematically consistent with
that with one electron per atom, as shown in Fig. 1 (a).

Although the following calculations were performed by Brinkman and
Rice, the calculations are applied to the effective mass. In the
case of Fig. 1 (c), by applying Eq. (9) to Eq. (6) and by
minimizing it with respect to $\bar\nu$, the number of the doubly
occupied atoms is obtained as
\begin{eqnarray}
\bar\nu=\frac{1}{4}(1+\frac{U}{8\bar\epsilon})&=&\frac{1}{4}(1-\frac{U}{U_c}),
\nonumber
\\ &=&\frac{1}{4}(1-{\kappa}{\rho}^2),
\end{eqnarray}
where $U_c$=8${\vert\bar\epsilon\vert}$ because of
$\bar\epsilon={\bar\epsilon_{\uparrow}}+{\bar\epsilon_{\downarrow}}<$0,
$U={\rho}^2U^{\prime}$ and $U^{\prime}={\kappa}U_c$.
$0<\kappa{\le}1$ is the correlation strength. By applying Eq. (10)
to Eq. (9) again, the effective mass is given by
\begin{eqnarray}
q_{\sigma}^{-1}=\frac{m^*}{m}&=&\frac{1}{1-(\frac{U}{U_c})^2},
\nonumber
\\ &=&\frac{1}{1-{\kappa}^2{\rho}^4}.
\end{eqnarray}

Although the separate conditions are $0<{\rho}{\le}1$  and
$0<{\kappa}{\le}1$, $m^{\ast}$ is defined under the combined
condition $0<{\kappa}{\rho}^2<1$ and is an average of the true
effective mass in the BR picture for metal phase (region C in Fig.
1 (b)). Eq. (11) is shown in Fig. 2(a). The effective mass
increases as it approaches ${\kappa}$=1 and ${\rho}$=1. For
${\kappa}{\ne}$0 and ${\rho}{\rightarrow}$0, the effective mass
decreases  and, finally, the metallic system undergoes a normal
(or band-type) metal-insulator transition; this transition is
continus. The system at ${\kappa}{\rho}^2=1$ can be regarded as
the insulating state which is the paramagnetic insulator because
${\bar{\nu}}=0$. At a $\kappa$ value (not one), the MIT from a
metal at a $\rho$ value of just below $\rho$=1 to the insulator at
both $\rho$=1 and $\kappa$=1 is the first-order transition on band
filling. This has been called the Mott transition by many
scientists including author. However, this is theoretically not
the Mott transition which is an first-order transition from a
value of $U$ to $U_c$ in a metal with the electronic structure of
one electron per atom at $\rho$=1, as given in the BR picture.
The Mott transition does not occur in real crystals because a
perfect single-phase metal with $\rho$=1 is not made. Conversely,
by hole doping (or electron doping to a metallic system with hole
carriers) of a very low concentration, the first-order transition
from the insulator with $\bar{\nu}$=0 to a metal can be
interpreted as an abrupt breakdown of the balanced critical
Coulomb interaction, $U_c$, between electrons. Then, the $U_c$
value in the insulator reduces to a $U$ value in a metal phase
and an insulating phase produces due to doping of opposite
charges, as shown in Fig. 1(d). This first-order transition with
band filling is very important result found in this picture,
which differs from the continuous (Mott-Hubbard or second-order)
transition by a large $U$ given by the Hubbard theory.

In order to obtain the expectation value of the energy in the
(paramagnetic) ground state, Eqs. (10) and (11) are applied to Eq.
(6). $E_g$ is given by
\begin{eqnarray}
E_g/l={\bar{\epsilon}}(1-{\kappa}{\rho}^2)^2.
\end{eqnarray}
As $U/U_c={\kappa}{\rho}^2$ approaches one, $E_g$ goes to zero.

In addition, the spin susceptibility in the BR picture is
 replaced by
\begin{eqnarray}
\chi_s&=&{\frac{m^{\ast}}{m}}{\frac{\mu_B^2N(0)}{(1-{\frac{1}{2}}N(0){\kappa\rho}^2U_c\frac{1+\frac{1}{2}
{\kappa\rho^2}}{(1+{\kappa\rho}^2)^2})}},
\end{eqnarray}
where $N(0)$ is the density of states at the Fermi surface and
$\mu_B$ is the Bohr magneton. The susceptibility is proportional
to the effective mass which allows the enhancement of $\chi_s$.

In such a case, $m^{\ast}$, $E_g/l$ and $\chi_s$ are the averages
of true effective values in region C in Fig. 1 (b), which is
described by the BR picture, and are justified only by means of
measurement. The true effective values in the BR picture are not
measured experimentally, as mentioned previously. In particular,
the magnitude of the true effective mass in the BR picture has the
same value regardless of the extent of region C, while the
measured effective mass depends upon the extent of region C.
Furthermore, validity of the BR picture and the Mott transition
was also discussed by theorists$[42,43]$, which is because the BR
picture has not been proved experimentally. However, the validity
is indirectly proved through the above picture.

\begin{figure}[t]
\vspace{0.1cm} \centerline{\epsfxsize=9.0cm\epsfbox{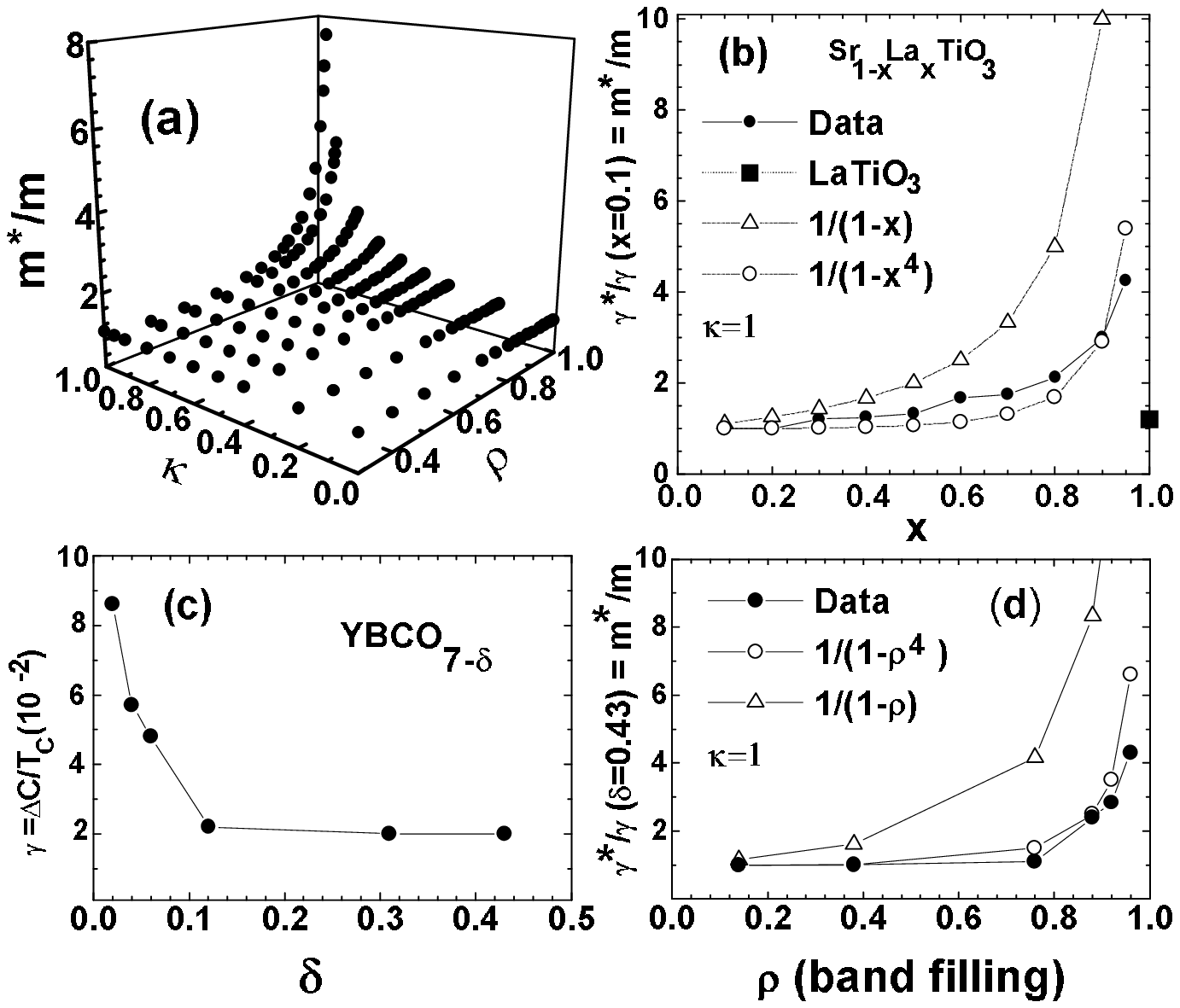}}
\vspace{0.1cm} \caption{\small{(a) The   effective mass,  $\frac
{m^{\ast}}{m}=\frac {1}{1-{\kappa}^2{\rho}^4}$. (b) Experimental
data (${\bullet}$) of the heat capacity  for
Sr$_{1-x}$La$_x$TiO$_3$ presented   by Tokura$[8]$ and
Kumagai$[9]$,    and  a   comparison   between $\frac
{m^{\ast}}{m}=\frac {1}{1-x^4}$ ($\circ$) and $\frac
{m^{\ast}}{m}=\frac {1}{1-x}$  calculated from the $t-J$
model$[5]$ ($\triangle$) and the Hubbard model$[6]$ on a square
lattice. Here, ${\rho}$ corresponds  to $x$ because the number of
quasiparticles with $x$ agrees  with   that of quasiparticles
obtained from the Hall coefficient$[8]$. Here, ${\kappa}$=1 and
$0<{\kappa}{\rho}^2<1$. (c) The heat-capacity coefficient
(${\triangle}c/T_c$) obtained by magnetic measurements for
YBa$_2$Cu$_3$O$_{7-\delta}$ superconductors by
Da{\"u}mbling$[48]$. (d) In cases where ${\delta}$=0 and
${\delta}$=0.45, the band is assumed to be full where ${\rho}$=1
and empty where  ${\rho}$=0, respectively. Thus
${\rho}$=1-${\delta}$/0.45. Here, ${\kappa}$=1 and
$0<{\kappa}{\rho}^2<1$. The above data were cited in reference
32.}}
\end{figure}

\section{Application to strongly correlated metals}
{\hspace{0.6cm}}This section describes the application of the
effective mass to experimental data of strongly correlated
metals. Sr$_{1-x}$La$_x$TiO$_3$ may be one of the most desirable
systems in  which the critical dependence of physical parameters
on the band filling can be investigated in the vicinity of the
MIT. For Sr$_{1-x}$La$_x$TiO$_3$, the number of 3$d$ electrons
varies from 0 to  1  with a   change of $x$ from SrTiO$_3$  to
LaTiO$_3$. The final  compound LaTiO$_3$ is  transformed to a
metallic   state  with a slight increase in $x{\geq}$0.05. Tokura
et  al.$[8]$ and Kumagai et al.$[9]$ measured the resistivity,
the magnetic susceptibility, the Hall coefficient $R_H$, and the
low-temperature heat  capacity as a function of $x$ for
Sr$_{1-x}$La$_x$TiO$_3$. The absolute value of $R_H^{-1}$
increases linearly with $x$ up to at  least $x$=0.95. The band
filling ${\rho}$  corresponds to $x$ because the number of doped
carriers with  $x$ agrees with the data for carriers obtained from
the Hall coefficient. The heat capacity in Fig. 3 of reference 8
is replotted, as shown in Fig. 2 (b). The effective mass of  Eq.
(11) closely follows the heat-capacity data where ${\kappa}$=1.
The heat capacity increases with $x$ and peaks at $x$=0.95. The
first-order transition is found between $x$=0.95 and $x$=1, which
is not the Mott transition defined by theory but the transition
with band filling. Tokura et al.$[8]$ and Kumagai et al.$[9]$
suggested that the increase of the heat capacity with $x$ is due
to  an enhancement of  the effective mass of correlated
electrons, although the carrier density increases with $x$;  the
coefficient, $\gamma$, of the heat capacity is proportional to
the effective mass in the Fermi Liquid theory. Moreover, the
large effective mass was also observed by optical methods$[10]$
and nuclear-magnetic-resonance experiments$[11]$.

Where $x<$0.95, $\kappa$=1 in Fig. 2(b) indicates that the true
effective mass has a constant value even though the extent of
region C in Fig. 1 (b) changes. The increase of $x$ up to $x$=0.95
corresponds to the increase of region C. Because region C is
described by the BR picture, the correlation strength of the BR
picture, $\kappa_{BR}$, can be found. When it is assumed that the
extent of the metal phase (region C) at $x$=0.95 is approximately
the same as that at $x$=$\rho$=1, a value of
$m^{\ast}/m=1/(1-{\rho}^4)$ at ${\rho}$=0.95 and $\kappa$=1 is
approximately equal to that of $m^{\ast}/m=1/(1-{\kappa}_{BR}^2)$
in the BR picture. Thus, ${\kappa}_{BR}$ = (0.95)$^2$=0.90 is
obtained, which indicates that La$_{1-x}$Sr$_x$TiO$_3$ is very
strongly correlated. Moreover, the decrease of the effective mass
from $x$=0.95 to $x$=0 is not the true effect, but the effect of
measurement.

For another correlated metal, Y$_{1-x}$Ca$_x$TiO$_3$, Kumagai et
al.$[9]$ obtained the same result as for Sr$_{1-x}$La$_x$TiO$_3$.
In V$_{2-y}$O$_3$, Carter et al.$[16]$ suggested that the
effective mass  may diverge at the Mott MIT under pressure and
decrease as the MIT is approached with $y$. The MIT with $y$
corresponds exactly to a band-type MIT in this extended BR
picture. McWhan et al.$[17]$ observed the sharp Mott MIT in
Cr-doped V$_2$O$_3$ at room temperature as functions of both Cr
concentration and pressure. The sharp Mott MIT is not the Mott
transition in the BR picture, but the first-order transition with
band filling $\rho$ in the extended BR picture. As another
experimental result, Khan et al.$[44]$ observed the first-order
transition at $x$=0.02 for $h$-BaNb$_x$Ti$_{1-x}$O$_3$, which may
also be in this context.

Morikawa et al.$[12]$ measured the decrease of the spectrum
intensity in the photoemission spectrum in  going from SrVO$_3$ to
CaVO$_3$, which corresponds to an increase in $U/W$ on the basis
of the Hubbard models. Here the increase of $U/W$ indicates an
increase of $\kappa$ with $\rho=1$, using the notations of the
extended BR picture. The decrease of the spectrum intensity is
different from the spectral function of the $d={\infty}$ Hubbard
model$[4]$, which suggests band narrowing rather than a
decreasing spectrum. Thus, they concluded that the effect of
long-range Coulomb interaction, which limits the mass enhancement
or  band narrowing near the Mott transition, is much larger  than
that of short-range interactions. However, in this extended BR
picture, the decrease of the spectrum intensity is interpreted as
a decrease of $U$ due to decreasing ${\rho}$ rather than
increasing $U/W$ (=$\kappa$, $\rho=1$). Thus, the effective mass
of quasiparticles decreases in going from SrVO$_3$ to CaVO$_3$,
which indicates that region C in Fig. 1 (b) gets small. The
transition with $x$ for Ca$_{1-x}$Sr$_x$VO$_3$ more closely
resembles a band-type MIT than the Mott MIT although the
theoretical analysis$[4]$ regards the transition as  a Mott
transition, because any behavior indicative of a first-order
transition due to a large $U$ is not found. The difference
between the extended BR picture and the $d=\infty$ Hubbard model
is discussed in Physic C or cond-mat/0001008 in reference 32.

Inoue et al.$[13]$ and  Makino et al.$[14]$ tried to control the
bandwidth W, assuming that the number of quasiparticles does not
vary with $x$  for single  crystals of Ca$_{1-x}$Sr$_x$VO$_3$.
However, investigations  of  the  BR feature near  the   Mott
transition showed  that the bandwidth did not change$[12]$.
Instead,  for high-resolution photoemission spectra, the spectral
intensity corresponding to quasiparticles decreased near the
Fermi energy$[12]$. This differs from the assumption of Inoue et
al. and Makino et al..

Ahn et al.$[45]$ investigated optical properties of (Ca,
Sr)RuO$_3$ films on the borderline of metal-insulator transition,
on the basis of the dynamical mean field theory(DMFT)$[46]$
regarded as the $d=\infty$ Hubbard mode. The effective mass of
quasiparticles increased with Sr concentration, although
CaRuO$_3$ is more metallic than SrRuO$_3$. This is the same as
the analysis for Ca$_{1-x}$Sr$_x$VO$_3$, but is different from
that in the extended BR picture. Unlike in the BR picture, the
number of quasiparticles is not conserved in the conduction band
as $U$ increases in the DMFT.

\section{Application to high-$T_c$ superconductors}
{\hspace{0.6cm}}This section describes the application of the
extended BR picture to cuprate oxide superconductors known to be
strongly correlated metals. Hongshun$[47]$, Da{\"u}mbling$[48]$,
W{\"u}hl$[49]$, and Loram$[35]$ observed the coefficient,
$\gamma$, of the specific heat capacity, which is proportional to
the effective mass corresponding to the 2D-DOS
(=$\frac{m^{\ast}}{\pi\hbar^2}$). The DOS showed a  jump in the
specific heat capacity near optimum doping. Eq. (11) seems in
line with Da{\"u}mbling's data, as shown in Figs. 2 (c) and (d).
The effective mass  decreases with decreasing ${\rho}$, which
approaches  a band-type  MIT. Near ${\rho}$=1, a behavior of
singularity is  shown  in Fig. 2 (d). Although  it is difficult
to  confirm whether the YBa$_2$Cu$_3$O$_{7-\delta}$(YBCO) of
${\delta}$=0 at ${\rho}$=1 is an insulator because the effective
mass near ${\rho}$=1 is divergent, the divergence is regarded as
the first-order MIT with band filling in Eq. (11).

\begin{figure}[t]
\vspace{0.1cm}
\centerline{\epsfxsize=10.0cm\epsfbox{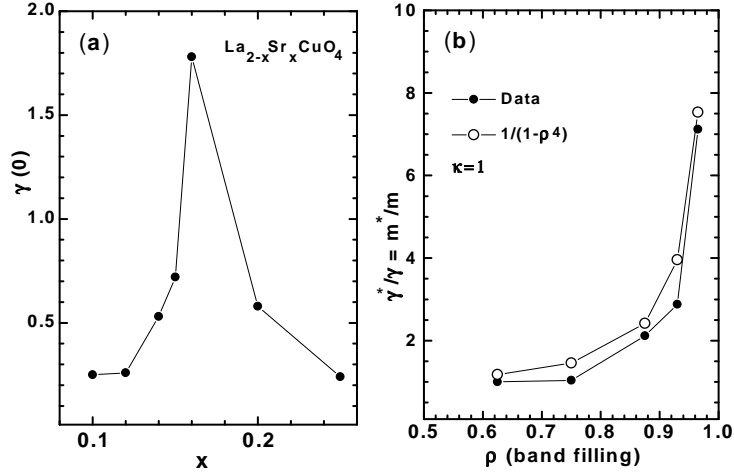}}
\vspace{0.1cm} \caption{\small{(a). Experimental data     of the
heat-capacity     coefficient for La$_{2-x}$Sr$_x$CuO$_4$
superconductors  presented  by Y. Hongshun$[47]$. (b). The
$m^{\ast}/m_{x=0.1}$ versus  ${\rho}$ is obtained from the
heat-capacity coefficient value to ${x=0.1}$  and an optimum
doping of $x$=0.16. At $x$=0.16, the  band is assumed to be half
full with ${\rho}$=1. Thus, ${\rho}$=$x$/0.16.  $\gamma$ is the
value of the heat capacity at $x$=0.1. The experimental data
(${\bullet}$) and $\frac {m^{\ast}}{m}=\frac {1}{1-{\rho}^4}$,
(${\circ}$), are shown. At $x$=0.15,  the effective  mass   is
estimated   to be $m^{\ast}/m{\approx}7.5$. Here, ${\kappa}$=1
and $0<{\kappa}{\rho}^2<1$.}}
\end{figure}

As for Loram's specific heat data of YBCO$_{6+x}$ in Fig. 3 and 4
of the reference 35, the anomaly step heights,
${\delta\gamma}(T_c)$, agree approximately with values estimated
by Da{\"u}mbling and W{\"u}hl. The anomalies near $x=0.57$ and
for ${x>}$0.92 have magnitudes of
${\delta\gamma}(T_c)/{\gamma_n\approx}$0.5 and
${\delta\gamma}(T_c)/{\gamma_n\approx}$2.5, respectively, which is
closely consistent with the values observed by Da{\"u}mbling. The
large values of $T_c$ and ${\delta\gamma}(T_c)$ with $x$ at
${x>}$0.9 showed little change. Fully oxygenated YBCO$_{7}$ was
slightly overdoped, which indicates that a crystal of YBCO$_{7}$
cannot be made, and that there is an instability similar to a
divergence at half filling. The instability seems related to the
metal-insulator instability$[34]$ and the divergence of the
effective mass near the transition from ${\rho\approx}$1 (not
one) to ${\rho}$=1. The physical meaning of the instability
indicates the transition from $\rho\neq$1 (or inhomogeneous
metalic phase or Fig. 1(b)) to $\rho$=1 (perfect homogeneous
metal or Fig. 1(a)), which does not indicate the van Hove
singularity. Thus, the data may be explained by the extended BR
picture.

In order to evaluate the correlation strength, $\kappa_{BR}$, in
the BR picture according to the same calculation method as used
for La$_{1-x}$Sr$_x$TiO$_3$ in a previous section, when it is
assumed that the extent of the metal phase (corresponding to
region C) at $\rho$=0.96 and $\kappa$=1 for YBCO$_{7-\delta}$ of
${\delta}{\approx}$0.04 is the same as that at ${\rho}$=1,
$m^{\ast}/m=1/(1-{\rho}^4)$ at ${\rho}$=0.96 is approximately
equal to $m^{\ast}/m=1/(1-{\kappa}_{BR}^2)$ in the BR picture.
Then ${\kappa}_{BR}={\rho}^2=(0.96)^2$=0.92 is obtained, which
indicates that the YBCO superconductor is strongly correlated.

In the case of La$_{2-x}$Sr$_x$CuO$_4$(LSCO) superconductors, the
heat capacity data are applied to Eq. (11), as shown in Fig. (3).
Band filling at $x$=0.15 is evaluated as $\rho$=0.96. By the same
calculation method, ${\kappa}_{BR}={\rho}^2$=0.92 is determined.
Moreover, the $\rho$ dependency of the effective masses, as shown
in Figs. 2 and 3, is the effect of measurement. Values of the true
effective masses are constant, though $\rho$ varies. The 2D-DOS
using the effective mass in Eq. (11) differs from an enhanced
2D-DOS combining both the van Hove singularity and the effect  of
mass enhancement$[15]$. The enhanced 2D-DOS did not agree with
experimental data of the heat capacity  at optimal doping $[15]$.
In addition,  the effective  mass, as measured by the de Haas-van
Alphen   effect at 2.3 K for YBa$_2$Cu$_3$O$_{6.97}$, was found
to be $m^*/m$=2.8-4.4$[50]$. The cyclotron mass of the organic
superconductor, (BEDT-TTF)$_2$Cu(NCS)$_2$, was found to be
$m_c/m_0{\approx}3.5[51,52]$. These large masses might be
attributed to the strong correlation in the BR picture. It is
suggested here that they may well be the cause of high-$T_c$
values for high-$T_c$ superconductors.

To account for high-$T_c$ superconductivity, the van Hove
scenario or singularity (vHs) has been introduced. The extended
saddle point near the Fermi energy has also been observed by
ultra-high-energy angle-resolved photoemission
spectroscopy$[19,20,24,25]$. It has been suggested that a $T_c$
value is dependent on the extent (or width and intensity of
spectra) of the flat band$[53]$ at the extended saddle-point
energy$[19,20,53]$. That is, the extent of the flat band,
corresponding to the number of quasiparticles in the flat band,
increases with increasing $T_c$. However, if this interpretation
is correct, it conflicts with the presence of quasiparticles at
the Fermi surface, as mentioned in the BR and extended-BR pictures
and the heat capacity data mentioned previously. Therefore, the
singularity in the effective mass is different from the extended
vHs, which may be a power law as a function of the difference
between the extended saddle-point energy and the Fermi energy.

A new Fermi-surface topology for the Bi-2212 system has been
suggested, instead of the flat band with the hole-like Fermi
surface$[54,55]$. In the new topology, a main band is centered
around ${\Gamma}$ with a smaller saddle point at ${\overline{M}}$
above $E_F$. It is described by an electron-like Fermi-surface
topology different from that of the flat band. The main-band
dispersion has been observed at the photon energy level of 33 $eV$
for Bi-2212 crystals. At $E_F$, the single-particle spectral
weight, $A(k,E_F)$, is at a peak corresponding to quasiparticles.
The peak can be regarded as band filling in the effective mass as
2D-DOS. Therefore, the extent of the flat band dependent on $T_c$,
as mentioned previously, can be interpreted as the magnitude of
the weight at the Fermi energy and as the extent of region C, too.
Furthermore, the new topology, such as the BR picture with
quasiparticles at the Fermi surface, is within the
Fermi-liquid-theory framework.

\section{Conclusion}
{\hspace{0.6cm}}The first-order transition on band filling for
Sr$_{1-x}$La$_x$TiO$_3$ (SLTO) and the experimental data of the
heat capacities, $\gamma^{\ast}/\gamma$=$m^{\ast}/m$, for SLTO,
YBCO$_{7-\delta}$, and LSCO were well correlated by the extended
BR picture. The true correlation strengths in the BR picture in a
metal phase (region C) were evaluated to be
0.90$\leq{\kappa}_{BR}<$1 for SLTO, 0.92$\leq{\kappa}_{BR}<$1 for
YBCO, and 0.92$\leq{\kappa}_{BR}<$1 for LSCO. The true effective
mass always has a constant value regardless of the magnitude of
$\rho$. High-$T_c$ superconductivity is attributed to the true
effective mass (regarded as the density of states) caused by the
large ${\kappa}_{BR}$ value.

The $\rho$ dependence of the effective mass is not an intrinsic
effect but one of measurement. In the general phase diagram, the
parabolic-$T_c$ curve and the declined linear of the pseudogaps
are also the effect of measurement. For high-$T_c$
superconductors, the MIT in the underdoped regime can be explained
without the concept of spin interaction. Gabovich's review paper
$[38]$ may give suggestions regarding the MIT based on the density
wave. Furthermore, the concept of measurement being correlated to
the extended BR picture can also be applied to explain the
measured physical properties of other synthetic metals or
insulators.
\\ \\
\Large{\bf Acknowledgements}\\ \\
%\vspace{0.2cm}
\normalsize{{\hspace{0.6cm}} I acknowledge Dr. Kwang-Yong Kang
for providing the research environment for this research. The use
of Fig. 3(a) was permitted
by Elsevier Science Co..}\\ \\
\Large{\bf References}
\\ \\
\small{$\noindent [1]$ N. F. Mott,
Metal-Insulator Transitions Chapter
3, (Taylor{\&}  Frances, 2nd edition,{\par}1990).\\
$[2]$ J. Hubbard, Proc. R. Soc. London A 276, 238 (1963);277, 237
(1964); 281, 401 (1964).\\
$[3]$ W. F. Brinkman and T. M. Rice, Phys. Rev. B{\bf 2}, 4302
(1970).\\
$[4]$ X. Y. Zhang, M. J. Rozenberg,  and G. Kotliar, Phys. Rev.
Lett.  {\bf  70}, 1666 (1993).\\
$[5]$ M. Kawakami and S. K. Yang, Phys. Rev. Lett. {\bf 65}, 2309
(1990).\\
$[6]$ N. Furukawa and M. Imada, J. Phys. Soc. Jpn. {\bf 60}, 3604
(1991).\\
$[7]$ M. C. Gutzwiller, Phys. Rev. 137, A1726 (1965).\\
$[8]$ Y. Tokura, Y. Taguchi, Y. Okada, Y. Fujishima, T. Arima, K.
Kumagai, and Y. Iye, {\par}Phys. Rev. Lett. {\bf 70}, 2126 (1993).\\
$[9]$ K. Kumagai, T. Suzuki, Y. Taguchi, Y. Okada, Y. Fujishima,
and Y. Tokura, Phys. {\par}Rev. B {\bf 48}, 7636 (1993).\\
$[10]$ Y. Fujishima, Y.  Tokura, T. Arima, and  S. Uchida, Phys.
Rev.  B {\bf 46}, 11167 (1992).\\
$[11]$ Y. Furukawa, I. Okamura, K. Kumagai, T. Goto, T. Fukase, Y.
Taguchi, and T. {\par}Tokura, Phys. Rev. B {\bf 59} 10550 (1999).\\
$[12]$ K. Morikawa, T. Mizokawa,  K. Kobayashi, A. Fujimori, H.
Eisaki, S. Uchida, F. Iga,{\par}and Y. Nishihara, Phys. Rev. B
{\bf 52}, 13 711 (1995).\\
$[13]$ I. H. Inoue, O. Goto, H. Makino, N. E. Hussey, and M.
Ishikawa, Phys. Rev. B {\bf 58},{\par}4372 (1998).\\
$[14]$  H. Makino, I.  H. Inoue, M.  J. Rozenberg, I.  Hase, Y.
Aiura,  and S. Onari, Phys.{\par}Rev. B {\bf 58}, 4384 (1998).\\
$[15]$ C. C. Tsuei, C. C. Chi, D. M.  Newns, P. C. Pattnaik, and
M. Daumling, Phys. Rev.{\par}Lett. {\bf 69}, 2134 (1992).\\
$[16]$  S. A. Carter, T.  F. Rosenbaum, P. Metcalf, J.  M. Honig,
and J. Spal, Phys. Rev. B {\par}{\bf 48}, 16841 (1993).\\
$[17]$ D. B. McWhan, T. M. Rice, and J. P. Remeika, Phys. Rev.
Lett. {\bf 23}, 1384 (1969).:D. {\par}B. McWhan, A. Menth,  J. P.
Remeika, W.  F. Brinkman, and T.  M. Rice, Phys. Rev. {\par}B
{\bf 7}, 1920 (1973).\\
$[18]$ N. Momono, M. Ido, T. Nakano, M. Oda, Y. Okajima, and K.
Yamaya, Physica C {\par}{\bf 233}, 395 (1994).\\
$[19]$ D. M. king, Z. X. Shen, D. S. Dessau, D. S. Marshall, C. H.
Park, W. E. Spicer, J. {\par}L. Peng, Z. Y. Li, and R. L. Greene,
Phys. Rev. Lett. {\bf 73}, 3298 (1994).\\
$[20]$ T. Yokoya, A. Chainani, T. Takahashi, H. Ding, J. C.
Campuzano, H. K. Yoshida, {\par}M. Kashi, and Y. Tokura, Phys.
Rev. B {\bf 54}, 13311 (1996).\\
$[21]$ J. L. Tallon and C. Bernhard, Phys. Rev. Lett. {\bf 75},
4552 (1995).\\
$[22]$ J. L. Tallon, G. V. Williams, C. Bernhard, D. M. Pooke, M.
P. Staines, J. D. {\par}Johnson, and R. H. Meinhold, Phys. Rev. B
{\bf 53}, R11972 (1996).\\
$[23]$ D. Y. Xing, M. Liu, and C. D. Gong, Phys. Rev. B, {\bf 44},
12525 (1991).\\
$[24]$ K. Gofron, J. C. Campuzano, A. A. Abrikosov, M. Lindroos,
A. Bansil, H. Ding, D. {\par}Koelling, and B. Dabrowski, Phys.
Rev. Lett. {\bf 73}, 3302 (1994).\\
$[25]$ J. Ma, C. Quitmann, R. J. Kelley, P. Almeras, H. Berger,
G. Margaritondo, {\par}and M. Onellion, Phys. Rev. B {\bf 51},
3832 (1995).\\
$[26]$ T. Timusk and B. Statt, ${\it ~Rep. ~Prog. ~Phys.}$ {\bf
62}, 61 (1999).\\
$[27]$ H. Ding, T. Yokaya, J. C. Campuzano, T. Takahasi, M.
Randeria, M. R. Norman, {\par}T. Mochiku, K. Kadowaki, and J.
Giapinzakis, ${\it Nature}$ {\bf 382}, 51 (1996).\\
$[28]$ S. H. Blanton, R. T. Collins, K. H. Kelleher, L. D. Rotter,
Z. Schlesinger, D. G. {\par}Hinks, and Y. Zheng, Phys. Rev. B
{\bf 47}, 996 (1993).\\
$[29]$ M. A. Karlow, S. L. Cooper, A. L. Kotz, M. V. Kelvin, P.
D. Han, and D. A. Payne, {\par}Phys. Rev. B {\bf 48}, 6499 (1993).\\
$[30]$ Hyun-Tak Kim, H. Uwe, and H. Minami, Advances in
Superconductivity VI {\par}(Springer-Verlag, Tokyo, 1994), P. 191.\\
$[31]$ J. L. Tallon and J. W. Loram, Physica C349, 53 (2001).\\
$[32]$ Hyun-Tak Kim, Physica C{\bf 341-348}, 259
(2000):cond-mat/0001008:cond-mat/0104055.\\
$[33]$ T. M. Rice and L. Sneddon, Phys. Rev. Lett. {\bf 47}, 689
(1981).\\
$[34]$ Hyun-Tak Kim, Phys. Rev. B {\bf 54}, 90 (1996).\\
$[35]$ J. W. Loram, K. A. Mirza, J. R. Cooper, W. Y. Liang, and
J. M. Wade, Journal of {\par}Superconductivity 7, 243 (1994).\\
$[36]$ J. W. Loram, J. L. Luo, J. R. Cooper, W. Y. Liang, and J.
L. Tallon, Physica {\par}C{\bf 341-348}, 831 (2000).\\
$[37]$ A. Junod, D. Eckert, T. Graf, G. Triscone, and J. Muller,
Physica C 162-164, 482 {\par}(1989).\\
$[38]$ A. M. Gabovich, A. I. Voitenko, J. F. Annett, and M.
Ausloos, Supercond. Sci. {\par}Technol. {\bf 14}, R1 (2001).\\
$[39]$ Tohr Ogawa, Kunihiko Kanda, and Takeo Matsubara, Prog.
Theor. Phys. {\bf 53}, {\par}(1975) 614.\\
$[40]$ Dieter Vollhardt, Rev. Mod. Phys. {\bf 56}, (1984) 99.\\
$[41]$ Patrick Fazekas, Lecture Notes on Electron Correlation and
Magnetism, Chapter 9, {\par}(World Scientific Co., 1999).\\
$[42]$ T. Moriya, BUTSURI (edited by Physical Society of Japan),
Vol 54, No. 1, (1999) {\par}48 (Japanese).\\
$[43]$ H. Fukuyama, M. Imada, and T. Moriya, BUTSURI {\par}(edited
by Physical Society of Japan), Vol 54, No. 2, (1999) 123
(Japanese).\\
$[44]$ M. N. Khan, Hyun-Tak Kim, H. Minami, and H. Uwe, Materials
Letters 47, 95{\par}(2001).\\
$[45]$ J. S. Ahn, J. Bak, H. S. Choi, T. W. Noh, J. E. Han, Yunkyu
Bang, J. H. Cho, and{\par}Q. X. Jia, Phys. Rev. Lett. {\bf
82}, 5321 (1999).\\
$[46]$ A. Georges, G. Kotliar, W. Krauth, and M. J. Rozenberg,
Rev. Mod. Phys. 68, 13 {\par}(1996).\\
$[47]$  Y. Hongshun,  Z. Xiaonong, Z.  Changfei, W.  Keqin, C.
Liezhao,  C. Zhaojia, Physica {\par}C{\bf 172}, 71 (1990).\\
$[48]$ Manfred Da{\"u}mbling, physica C {\bf 183},293 (1991).\\
$[49]$ H. W{\"u}hl, R. Benischke, M. Braun, B. Frank, O. Kraut,
R. Ahrens, G. Br{\"a}uchle, H. {\par}Claus, A. Erb, W. H. Fietz,
C. Meingast, G. M{\"u}ller-Vogt, and T. Wolf, Physica C {\par}{\bf
185-189}, 755 (1991).\\
$[50]$  J. L. Smith, C. M. Fowler, B. L. Freeman, W. L. Hults, J.
C. King, and F. M. {\par}Mueller, Advances in superconductivity
III,(Springer-Verlag, Tokyo,1991), p. 231.\\
$[51]$ N. Toyota, T. Sasaki, K. Murata, Y. Honda, M. Tokumoto, H.
Nando, N. Kinoshita, {\par}H. Anzai, T. Ishiguro, and Y. Muto, J.
Phys. Soc. Japan {\bf 57}, 2616 (1988).\\
$[52]$ K. Oshima, H. Urayama, H. Yamochi, and G. Saito, Physica C
{\bf 153-155}, 1148 {\par}(1988).\\
$[53]$ D. S. Dessau, Z.-X. Shen, D. M. King, D. S. Marshall, L. W.
Lombardo, P. H. {\par}Dickinson, A. G. Loeser, J. DiCarlo, C. H.
Park, A. Kapitulnik, and W. E. Spicer, {\par}Phys. Rev. Lett.,
{\bf 71}, 2781 (1993).\\
$[54]$ Y. D. Chuang, A. D. Gromko, D. S. Dessau, Y. Aiura, Y.
Yamaguchi, K. Oka, A. J. {\par}Arko, J. Joyce, H. Eisaki, S. I.
Uchida, K. Nakamura, and Yoichi Ando, Phys. Rev. {\par}Lett. {\bf
83}, 3717 (1999).\\
$[55]$ A. D. Gromko, Y. D. Chuang, D. S. Dessau, K. Nakamura, and
Yoichi Ando, {\par}cond-mat/0003017.}

\end{document}